\documentclass{article}
\usepackage[utf8]{inputenc}
\usepackage[round]{natbib}
\usepackage{physics}
\usepackage{lmodern}
\usepackage{geometry}
\usepackage[english]{babel}
\usepackage{url}

\usepackage{graphicx}
\title{Holography without holography: How to turn inter-representational into intra-theoretical relations in AdS/CFT} 

    \author{Rasmus Jaksland\thanks{rasmus.jaksland@ntnu.no, Department of Philosophy and Religious Studies, NTNU Norwegian University of Science and Technology, NTNU Dragvoll, 7491 Trondheim, Norway} \and Niels S. Linnemann\thanks{niels.linnemann@uni-bremen.de, Institute of Philosophy, University of Bremen, 28359 Bremen, Germany} }

\date{For the published version (open access), please see \url{https://doi.org/10.1016/j.shpsb.2020.04.007}}

\usepackage{amsmath}
\usepackage{amssymb}
\usepackage{amsfonts}

\begin{document}
\maketitle

\newpage
\begin{abstract}
We show by means of the AdS/CFT correspondence in the context of quantum gravity how inter-representational relations---loosely speaking relations among different equivalent representations of one and the same physics---can play out as a tool for intra-theoretical developments and thus boost theory development in the context of discovery. 
More precisely, we first show that,
as a duality, the AdS/CFT correspondence cannot in itself testify to the quantum origin of gravity (though it may be utilized for this purpose). 
We then establish through two case studies from emergent gravity (Jacobson (2016), Verlinde (2017)) that the holographic AdS/CFT correspondence can, however, still excel as a guiding principle towards the quantum origin of gravity (similar in nature to quantisation).
\end{abstract}

\tableofcontents

\section{Introduction}
Ever since its discovery, the AdS/CFT\footnote{\textit{AdS} stands for Anti-de-Sitter, \textit{CFT} for conformal field theory.} correspondence has intrigued researchers in many areas of physics. The correspondence conjectures a duality between $D=10$ type IIB superstring theory on $AdS_5 \times S^5$ and $D=4$, $\mathcal{N}=4$ Super Yang-Mills theory defined on a fixed spacetime background conformal to the asymptotic boundary of $AdS_5$; here $D$ denotes spacetime dimensions, and $\mathcal{N}$ number of supersymmetries.\footnote{There are other examples of AdS/CFT correspondences, e.g. $AdS_4/CFT_3$ (involving type IIA superstring theory) \citep{aharony__2008} and $AdS_3/CFT_2$ (involving type IIB superstring theory) \citep[section 5]{maldacena_large-n_1999}. Our focus here shall be on $AdS_5/CFT_4$, but all said applies equally well to other AdS/CFT correspondences. The same goes for other examples of gauge/gravity dualities, for instance the one from heterotic string theory \citep{chen_gauge/gravity_2013}} 
The AdS/CFT correspondence is an example of a gauge/gravity duality: a duality between a $D$-dimensional gauge theory and a $D+1$-dimensional theory of gravity.
\footnote{While the AdS/CFT correspondence is a relation between a four-dimensional and a ten-dimensional theory, it is still a gauge/gravity duality. This is so since the ten dimensional spacetime includes five compact dimensions---those of the $S^5$---such that the boundary of the spacetime is the four-dimensional boundary of $AdS_5$.} 
As such, gauge/gravity dualities and therefore the AdS/CFT correspondence realise a holographic setting: the physics of the system may be represented both by a theory defined in a volume enclosed by a surface and by another theory defined on the surface enclosing the volume (the physics of the system can be seen as being projected from the boundary of the volume).

From the perspective of quantum gravity, the AdS/CFT correspondence is enticing: it promises a relation between a quantum field theory and a theory of gravity beyond semi-classical gravity and effective field theories, i.e. an apparent breakthrough in the attempt to unify quantum degrees of freedom with gravitational ones. In the first part of this paper we argue that the AdS/CFT correspondence, qua duality, cannot immediately fulfill this promise to be a general guide to the quantum origin of gravity, by which we mean a general prescription of how classical gravity emerges from an underlying quantum level of description. We argue that the only theory of quantum gravity \textit{directly} advanced by the AdS/CFT correspondence is the type of string theory found on the AdS side; typically type IIB superstring theory. In involving a duality, any entry of the AdS/CFT dictionary is a statement about the relation among quantities in two different equivalent representations of the same underlying bare theory.\footnote{See the next section for a precise clarification.} Thus, when the AdS/CFT correspondence relates gravitational degrees of freedom to quantum degrees of freedom on the boundary, this is not implying that gravity emerges from boundary degrees of freedom, just as the validity of Fourier transformation is not demonstrating that momentum degrees of freedom emerges from positional ones, or \textit{vice versa}. 
What we seek in a theory of quantum gravity is not a re-representation of a theory in new guises (in principle, if you know one representation of the bare theory, you do not learn anything extra about it through another representation). Rather, what we are after are the underlying degrees of freedom from which general relativity and the standard model are expected to emerge. 
This is not what the AdS/CFT correspondence provides although it does serve to improve our understanding of the theory of quantum gravity we find on the AdS side namely string theory (more on this is section \ref{section:31}).

The AdS/CFT correspondence, however, can nevertheless be of general utility in the search for theories of quantum theory: 
In the second (constructive) part of this paper, we will defend the view that the discoveries of the AdS/CFT correspondence---in order to allow for general advances towards the quantum origin of gravity---must and can be ``activated" in a non-trivial manner. By this we will mean ways in which to use the intriguing insights of the AdS/CFT correspondence---especially as regards to the relation between quantum and gravitational degrees of freedom---as something more than the mere re-representations they are at the outset. This will involve using the AdS/CFT correspondence as a heuristic as proposed by \citet{de_haro_heuristic_2018}. As will be argued, this cannot leave the correspondence intact and the methodology therefore involves breaking the duality to achieve emergence; thus echoing \citet{teh_holography_2013}, \citet{dieks_emergence_2015}, and \citet{de_haro_dualities_2017} among others. Our treatment, however, differs somewhat from these in that it diagnoses the tension between duality and emergence as one between inter-representational and intra-theoretical relations. While this might look like a mere linguistic relabelling, it serves to signify an important obstacle in the attempt to relieve this tension: 
Exploiting the duality in the search for the quantum origin of gravity cannot simply involve the stipulation that the relations coming out of the AdS/CFT correspondence are not exact.\footnote{Following \citet{de_haro_heuristic_2018}, dualities are inexact if ``they are not instantiated by the models in an exact manner." (footnote 19)} Rather, the duality relations of interest have to be changed individually from inter-representational relations into intra-theoretical ones by embedding the dual elements into the same description of reality.
The ambition for the second part of the paper is to provide a more detailed procedure for how to achieve this---the methodology we denote `holography without holography'---than those already found in the literature.

In a nutshell, this methodology amounts to a procedure whereby symmetric duality relations of the AdS/CFT correspondence serve as guiding principles in the search for the relation between underlying quantum degrees of freedom and gravity \textit{without}, however, realising the AdS/CFT correspondence and holography in particular. The methodology has already been implicitly implemented in works by \citet{jacobson_entanglement_2016} and \citet{verlinde_emergent_2017} who in their respective ways use insights from the (holographic) AdS/CFT correspondence in an exploration of the emergence of gravity within a \textit{non-holographic} context: The work of \citet{jacobson_entanglement_2016} suggests taking the CFT degrees of freedom as an inspiration (but nothing more!) for the degrees of freedom from which gravity arises. Relations between these underlying degrees of freedom and gravity that resemble those of the AdS/CFT correspondence are then proposed to hold in virtue of thermodynamic effects rather than as holographic duality relations. In this way, what used to be different representations of the same bare theory are---modulo proper modifications---turned into thermodynamic relations of coarse-graining within the same representation. Verlinde's emergent gravity program (\citeyear{verlinde_emergent_2017}) also uses holography as an inspirational tool for an actual intra-theoretical statement from an inter-representational statement but does not completely leave it to this: In his account of de Sitter space as an excited state of Anti-de Sitter space (thus the ground state), holography still genuinely features at the ground state level.

Before we proceed a brief terminological remark is in order. We will refer to relations that obtain between different dual representations as inter-representational relations. Equalities between quantities of different dual representations, i.e. quantities standing in an inter-representational relation, will be marked with the superscript `dual': `$\stackrel{dual}{=}$'. In contrast, the mere equality sign `$=$' will be used exclusively for statements of equality between properties of the same representation, thus intra-theoretical\footnote{We stick to the standard nomenclature here according to which dualities are relations between theories; that is, each representation is a theory. This way of talking, however, stands in mild terminological tension with thinking of a supposed common core underlying the two representations as the actual theory at play.},  as opposed to inter-representational relations. In this terminology, most relations encountered in physics are intra-theoretical relations: think of Newton's second law, the Schrödinger equation, the Bekenstein-Hawking formula. Inter-representational relations include, as mentioned, dualities, but a more mundane example are the Fourier transformations that relate time and frequency or space and momentum representations to one another (as familiar from electrodynamics and quantum mechanics) or change in basis/coordinates more generally. We will return to this distinction in the context of the AdS/CFT correspondence below.

In the following, we first give an introduction to the AdS/CFT correspondence and holography (section 2). We then argue how the AdS/CFT correspondence on its own does not directly provide general insights into the quantum origin of gravity (section 3). Instead, we work out, through two case studies, how the holographic AdS/CFT correspondence can star as an indirect methodological tool (section 4). We develop this further in section 5 and argue that the AdS/CFT correspondence can be employed as what we will call an analytic guiding principle in the context of quantum gravity analogous to quantisation.

On a more general level, the paper aims at illustrating how inter-representational relations---relations among different equivalent representations---can still play out as tools for intra-theoretical advances in the context of discovery.

\section{The AdS/CFT correspondence and holography} 
In this section, we give a short introduction to the AdS/CFT correspondence and point out its holographic nature. 

Naively, two theories are often said to be dual if they are equivalent with respect to all their physically/empirically significant elements.\footnote{Cf. \citet[62]{rickles_dual_2017}: ``a pair of theories is said to be dual when they generate the same physics."} 
Nevertheless, as for instance argued by \citet{deHaroButterfield2017}, theories can be dual even if they are not about the same physics/the same empirical content: in fact, two dual theories each of which are seen as embedded within different (external) physical contexts are generally not at all empirically equivalent.\footnote{\citet{deHaroButterfield2017} give the example of the Kramers-Wannier duality between the high and low temperature regimes of the statistical mechanics of a lattice.}
So, rather than linking dualities to a notion of physical or empirical equivalence, dualities should strictly speaking be conceived of as formal relations holding between uninterpreted theories, i.e. theories not linked to the world yet. We will thus follow the so-called 'Schema' for dualities as explicated in \citet{de_haro_spacetime_2017}, \citet{deHaroButterfield2017}, and \citet{de_haro_heuristic_2018}: two uninterpreted theories (which we shall denote `representations') are dual iff they are isomorphic representations of one and the same common core theory (called bare theory).\footnote{For the present purposes, we will disregard any subtleties arising from the multiple interpretive stances one may take towards the metaphysics of dualities (see \citet{read_interpretation_2016} and \citet{le_bihan_duality_2018} for an overview of these).} Each representation consists of a copy of the bare theory and some specific structure. This schema is straightforwardly illustrated through an analogy with group theory: an abstract group (the bare theory) can be represented by concrete group representations which can be thought of as a pair containing the structure linked to the bare theory and specific structure linked to its concrete, individual nature as a specific representation. The abstract group $SU(2)$ is for instance represented by various concrete matrix groups. These concrete group representations are of course not generally isomorphic to one another; only if they are isomorphic do they count as dual.

A simple example of two isomorphic group representations is the group of all real numbers with addition $G=(\mathbb{R},+)$ that is isomorphic to the positive real numbers with multiplication $G'=(\mathbb{R}^+ , \cdot )$ under the group isomorphism $f: G \rightarrow G', x \mapsto \exp{x}$: if $x+y=z$ then $\exp{x} \cdot \exp{y} = \exp{z}$ for all $x,y,z \in \mathbb{R}$. They are two different representations of the same abstract group. If one wants to add real numbers or multiply exponentials, then one can use either representation. As an illustration, consider Alice, a stubborn and mathematically gifted kid with a single shortcoming: she struggles with addition. Consequently, she has simply given up on addition, and instead adopted (1) the exponential sequence $\exp{1}, \exp{2}$, $...$ for counting, and (2) multiplication for adding them. To make things easier, she has memorized a type of dictionary for moving back and forth between her numbers, the exponentials, and the ordinary numbers used by everyone else where the entries of this dictionary are given by the isomorphic function $f(x) = \exp{x}$. With this system, Alice can do what everyone else does with ordinary numbers: she can make sure that she has coin enough to buy her three chosen pieces of candy, she can follow math classes, she can assess which lake in the park holds more swans and how many there are in total. All cases where we use real numbers and addition, Alice can use exponentials and multiplication, irrespective of what the numbers represent! 

As such, the story of Alice makes vivid the formal character of isomorphisms and thus dualities: they are independent of interpretation; an aspect that will become important in our criticism of the role of AdS/CFT correspondence in the search for the quantum origin of gravity. If some situation can be represented in terms of $(\mathbb{R},+)$, then it can be represented by $(\mathbb{R}^+ , \cdot )$. There is no additional question whether the isomorphism is realised; Alice does \textit{not} have to check in every case where real numbers are added, whether she can represent the situation using exponentials and multiplication. The formal relation, the isomorphism, guarantees that whenever people add (or subtract) real numbers, Alice can multiply exponentials instead. But this also entails that Alice's knowledge of the isomorphism is not an insight about that which is counted and added, but rather an insight about the abstract mathematical group instantiated by counting and adding. Generally, isomorphisms do not give us degrees of freedom that were not already there.\footnote{Thanks to an anonymous reviewer for pressing us here.} Knowing two representations is helpful if you, like Alice, struggle with one of them, but this does \textit{not} entail that one representation can disclose something absent in the other (though different aspects might be manifest). This is so even if we want to insist that there are two swans in the lake, and not $\exp{2}$ as Alice claims. In a sense, only $(\mathbb{R},+)$ represents when it comes to swans in lakes. But even this makes no difference for Alice! Even if $(\mathbb{R}^+ , \cdot )$ somehow misrepresent the ontology of the situation, Alice can still use it to find out how many swans the park's lakes have in total. Again, this is so because the isomorphism is a purely formal relation that is unaffected by the interpretation of either representation.

While such dualities are well known in mathematics, they are more surprising in physics. One of the most profound examples known from physics---due to pioneering work by Juan \citet{maldacena_large-n_1999}---is the conjectured\footnote{The AdS/CFT correspondence is strictly speaking just a conjecture, that is a so-far unproven theorem. For an overview on reasons why to accept the AdS/CFT correspondence nevertheless, see \citet[section 5.4]{WallaceStatisticalMechanics} and \citet[chapter 6-8]{ammon_gauge/gravity_2015}.}
duality between $D=10$ type IIB superstring theory on an asymptotically $AdS_5 \times S^5$ background (the AdS side) and $D=4$, $\mathcal{N}=4$ Super Yang-Mills theory (SYM) defined on a background identical to the conformal boundary of $AdS$---known as the AdS/CFT correspondence.\footnote{For the Poincaré patch of AdS spacetime, the conformal boundary is Minkowski spacetime, and for (the universal covering of) global AdS (in $D+1$ dimensions), the conformal boundary is $\mathbb{R} \times S^{D-1}$.} 
As a duality, the AdS/CFT correspondence conjectures an isomorphism between these two theories or rather, qua duality, these two representations. As was the case with the isomorphism between $(\mathbb{R},+)$ and $(\mathbb{R}^+ , \cdot )$, a dictionary exists that translates between the string theory with a dynamical spacetime on the AdS side to the quantum field theory without gravity on the CFT side, however remarkable this may seem. Any expression and operation on the AdS side can be translated to an equivalent expression and operation on the CFT side, and \textit{vice versa}, modulo technical complications analogous to Alice's potential problems relating to the evaluation and multiplication of exponentials. 

While the AdS/CFT correspondence in its general form involves a duality between string theory and conformal field theory, in the low energy, weak coupling limit\footnote{More precisely, this is the limit where the string length, $l_s$, is much smaller than the characteristic length scale of the spacetime background and the string coupling, $g_s$, is much smaller than one.} the AdS side can be approximated by type IIB supergravity from which one can derive the Einstein-Hilbert action with negative cosmological constant (together with some additional matter fields following from the limit from superstring theory\footnote{One essentially obtains supergravity upon inserting additional fields.}). Taking the corresponding limit on the CFT side,\footnote{On the CFT side, this is the limit where the rank of the gauge group goes to infinity and the 't Hooft coupling is large but finite. For further details, see \citet[chapter 5]{ammon_gauge/gravity_2015}.} one arrives at a duality between strongly coupled $\mathcal{N} = 4 $ SYM and semi-classical gravity. Thus, it promises to relate semi-classical gravity to quantum field theory; semi-classical gravitational degrees of freedom to quantum ones.

As its name indicates, the AdS/CFT correspondence is a correspondence and not merely a duality,\footnote{Thank you to an anonymous reviewer of this journal for pressing us on this issue.} though it is a duality nonetheless. Being a correspondence, this duality comes with the additional interpretative commitment that the two sides of the duality are indeed representations of the same physics, and not just an abstract structural similarity between two distinct physical systems (see \citet{deHaroButterfield2017} for more on this difference). This suggests an immediate utility of the AdS/CFT correspondence in overcoming difficulties in one representation by translating to the other; again Fourier transformations and their use in moving between position and momentum representation is a good illustration. We take no issue with this use of the AdS/CFT correspondence as a transformation. What we want to argue is that this use in itself will not attest to the quantum origin of gravity though it might be of service in the exploration of theories, such as type IIB string theory, that do so. 
The AdS/CFT correspondence has also been employed to explore issues of more conceptual nature relevant for quantum gravity research\footnote{A number of these were raised to our attention by an anonymous reviewer of this journal.} such as the information paradox (see \citet{harlow_Jerusalem_2016} for a review), cosmic censorship and cosmological bounce \citep{engelhardt_new_2016}, locality \citep{hamilton_holographic_2006}, and the relation between entanglement and geometry that we will discuss in more detail below. These, however, are still just exploiting the resources of re-representation. They translate from one side of the duality---typically the AdS side since it features gravity---to the other where different technical capacities can be used to explore the same (qua correspondence and not only duality) physics. From this procedure we can, for instance, learn about the quantum nature of AdS black holes. These lessons then promise to generalize to other black holes and thus quantum gravity in general under the \textit{additional} assumption that they do not rely on elements peculiar to this holographic setting. This additional assumption is not and cannot be sanctioned by the AdS/CFT correspondence, so even these conceptual insights, in principle, go beyond the direct import of the AdS/CFT correspondence. In fact, they are one typical example of the heuristic use of the AdS/CFT correspondence that by their straightforward use of re-representation are rather different from the heuristic function of the AdS/CFT correspondence developed in section 4 and 5.

\subsection{Holography}
Soon after its discovery, the AdS/CFT correspondence was linked to the notion of holography \citep{witten_anti_1998}.
Holography was first conceived by \citet{t_hooft_dimensional_1994} as the conjecture that
\begin{quote}
given any closed surface, we can represent all that happens inside it by degrees of freedom on this surface itself. This, one may argue, suggests that quantum gravity should be described entirely by a topological quantum field theory, in which all physical degrees of freedom can be projected onto the boundary \citep{t_hooft_dimensional_1994}\footnote{See also \citep{susskind_world_1995}.}.
\end{quote}
In other words, a system is holographic if and only if the physics of the system can be represented both by a theory defined in the volume enclosed by the surface---commonly referred to as the bulk---and by another theory (with different degrees of freedom) defined on a surface enclosing the volume---commonly referred to as the boundary.

Thus, holography as formulated by 't Hooft---a holographic duality---obtains if a system admits both a $D$-dimensional and a $(D+1)$-dimensional representation such that the $D$-dimensional representation is defined on a background identical to the boundary of the $(D+1)$-dimensional representation.

It may not be immediately clear how holography is realised by the AdS/CFT correspondence. After all, the AdS/CFT correspondence involves a duality between a four-dimensional gauge theory and a ten-dimensional theory of gravity. However, only five of these dimensions---those of $AdS_5$---are extended dimensions and therefore relevant from the point of view of holography. Hence, the bulk degrees of freedom can be encoded by boundary degrees of freedom as required by holography. According to the AdS/CFT correspondence, the five-dimensional theory of gravity has an alternative representation in terms of a four dimensional theory ``living" on its boundary.

\subsection{Gravity from Entanglement}\label{Gravity from Entanglement}
\begin{figure}
	\begin{center}
		\includegraphics[scale=0.3]{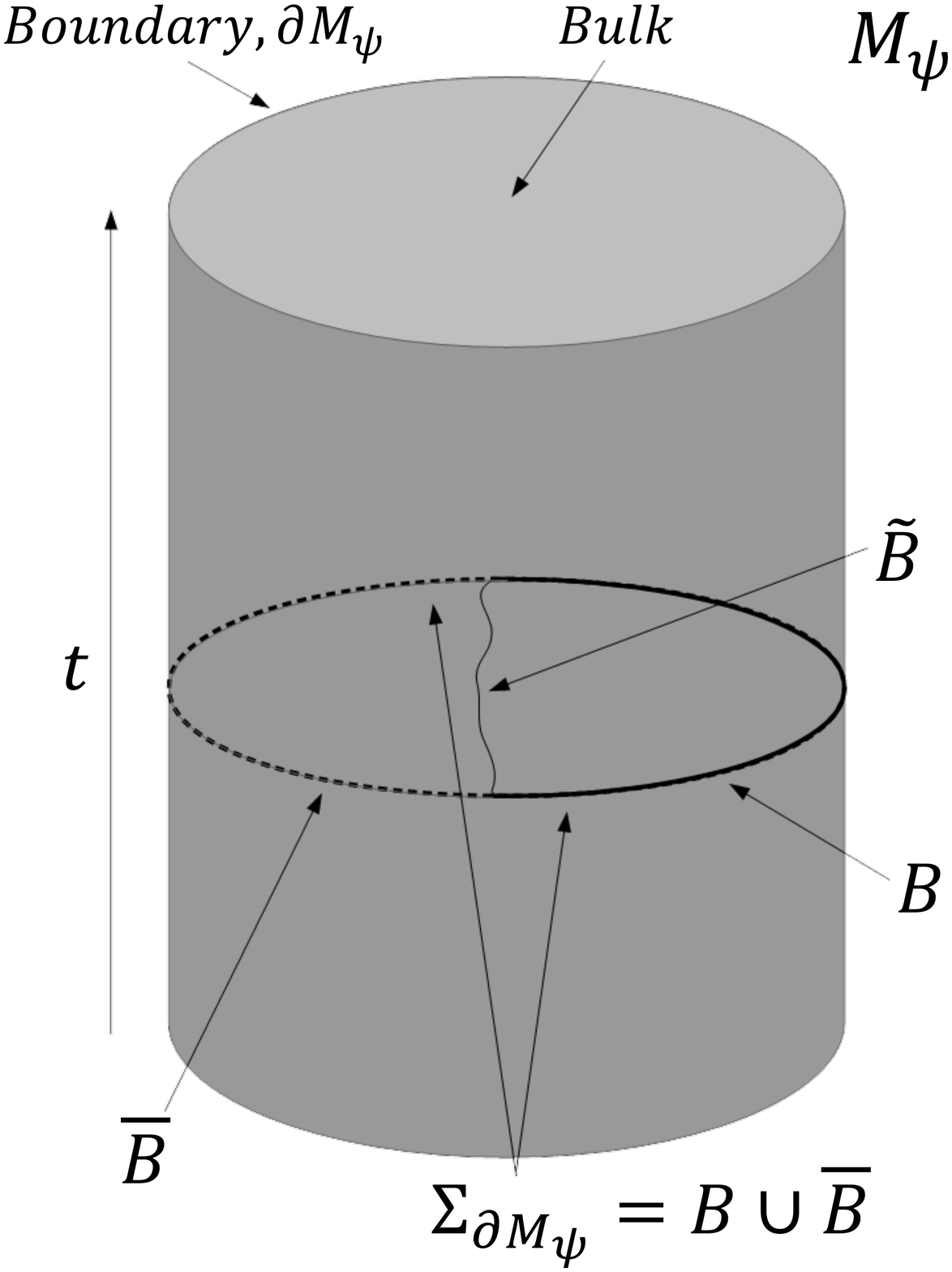}
		\caption{\label{cylinder} For the purpose of illustration, a spacetime is depicted here with boundary $S^{D-1} \times R$. With the AdS metric in the interior of the cylinder, this is a depiction of the universal covering of global AdS spacetime. Figure is taken from \citet{jaksland_entanglement_2020}.}
	\end{center}
\end{figure}

The understanding of the AdS/CFT correspondence has recently advanced with the discovery of a relation between the entanglement entropies for subsystems on the CFT side and the area of extremal co-dimension two-surfaces on the AdS side \citep{ryu_holographic_2006}. 
To introduce this, some set-up is required. Since this relation proves important in the discussion of \citet{jacobson_entanglement_2016}, we will present this setup more carefully than what might seem at first necessary: Consider a CFT state $\ket{\Psi}$ with an AdS dual that features a classical spacetime $M_{\Psi}$ (see the illustration on figure \ref{cylinder}). By holography, $\ket{\Psi}$ is a state in the Hilbert space for a CFT which itself is defined on a spacetime identical to the asymptotic boundary of $M_{\Psi}$ (denote it as $\partial M_\Psi$). To construct the Hilbert space of the CFT, define a spatial slice of $\partial M_\Psi$ (denote it as $\Sigma_{\partial M_\Psi}$). We then have $\ket{\Psi} \in \mathcal{H}_{\Sigma_{\partial M_{\Psi}}}$. Now, divide $\Sigma_{\partial M_\Psi}$ into two regions $B$ and $\overline{B}$, such that $B \cup \overline{B} = \Sigma_{\partial M_\Psi}$. We can regard the full quantum system as composed of two subsystems, $Q_B$ and $Q_{\overline{B}}$, associated with the two spatially separated regions $B$ and $\overline{B}$. As a consequence, the Hilbert space of the full system can be decomposed as a tensor product of the Hilbert spaces of $Q_B$ and $Q_{\overline{B}}$.

On the CFT side, we define the entanglement entropy $S_B = -\tr ( \rho_B \log ( \rho_B ) )$ with the density matrix $\rho_B = \tr_{\overline{B}} ( \ket{\Psi} \bra{\Psi})$. $S_B$ is the entanglement (von Neumann) entropy associated with entanglement between the quantum systems $Q_B$ and $Q_{\overline{B}}$ over an entangling surface that coincides with the boundary of $B$, that is $\partial B$. 

On the AdS side, we define $\tilde{B}$; the co-dimension two-surface of minimal area whose boundary (``endpoints") is such that it separates $B$ from $\overline{B}$, i.e. $\partial \tilde{B} \equiv \tilde{B} \vert_{\partial M_{\Psi}} = \partial B$ (see figure \ref{cylinder}). 
We are now in the position to formulate the mentioned relation between entanglement entropy for subsystems on the CFT side, and area of extremal co-dimension two-surfaces on the AdS side: the so-called Ryu-Takayanagi formula conjectures that

\begin{equation} \label{eq:RyuTakayanagi}
S_B \stackrel{dual}{=} \frac{A(\tilde{B})}{4 G_N \hbar}
\end{equation}
where $A(\tilde{B})$ is the area of $\tilde{B}$.

The Ryu-Takayanagi formula relates entanglement on the CFT side to a spacetime surface property---its area---on the AdS side. Thereby, the Ryu-Takayagani formula cashes out the AdS/CFT correspondence in a vivid manner by relating entanglement to spacetime. It also \textit{seems} to shed new light on the Bekenstein-Hawking formula that relates the horizon area, $A_{BH}$, to the entropy, $S_{BH}$, of stationary black hole. Now, AdS/CFT correspondence or not, the Bekenstein-Hawking formula is still expected to be satisfied in the bulk. So if $A(\tilde{B}) = A_{BH}$ then $S_{BH} \stackrel{dual}{=} S_{B}$ where $B$ is the full Cauchy surface on which the thermal CFT state is defined, i.e. the complement of $B$ is empty. Again, this is a relation among quantities of different representations. However, it cannot but make one speculate whether the origin of these degrees of freedom are related. Since the origin of $S_{B}$ is known to be the entanglement entropy of some CFT state---it might be tempting to envisage that the black hole degrees of freedom simply are these CFT degrees of freedom. This idea is by no means foreign in the physics literature: ``one can regard the origin of the black hole entropy as the entropy of the corresponding gauge theory, namely the number of microscopic states of the gauge theory" \citep[34-35]{natsuume_ads/cft_2015}.\footnote{Describing this as holography already alludes to this ontological prioritising of boundary since in a hologram the information to create a three dimensional image is stored in a two dimensional surface. It therefore appears three dimensional under certain circumstances, but it is really two-dimensional.} It is even often conjectured for general spacetimes in the context of the AdS/CFT correspondence: a ``spacetime in which gravity operates is emergent from the collective dynamics of the quantum fields; the latter by themselves reside on a rigid spacetime sans gravity" \citep[3]{rangamani_holographic_2017}.\footnote{Other examples are \citet[178]{horowitz_gauge/gravity_2009}, \citet[3]{faulkner_gravitation_2014}, and \citet[2]{hubeny_ads/cft_2015}.} So does this not suggest that AdS/CFT correspondence might provide a theory for the quantum origin of gravity? It is after all one of the central expectations of such a theory that it will feature the degrees of freedom that give rise to the Bekenstein-Hawking formula.

\section{AdS$\stackrel{dual}{=}$CFT and the quantum origin of gravity}\label{AdS/CFT}
In this section, we argue that 
the AdS/CFT correspondence \textit{qua involving a duality} can merely offer re-representations. It can, in other words, not directly show how gravitational degrees of freedom emerge from quantum ones despite its enticing relations between these and the (apparent) hopes to the contrary. The AdS/CFT correspondence cannot be a direct guide for this central component of a theory of quantum gravity.

Broadly speaking, we take the problem of quantum gravity to consist in improving on the lack of predictivity\footnote{Perturbative quantum gravity can be treated as an effective field theory and thus as effectively renormalizable only far below the Planck energy scale --- at higher energy scales, measurements need to be taken again and again on the coupling constants to make statements about that energy scale in question.} perturbative quantum gravity suffers from due to its non-renormalizability. The means of improvement here standardly include attempting non-perturbative quantisation instead (such as canonical quantum gravity, and canonical loop quantum gravity), taking the renormalization group flow seriously (asymptotic safety), and reproducing perturbative quantum gravity only in some lower energetic limit (such as done by string theory). Following \cite{CrowtherLinnemann}, we can conceive of the problem of quantum gravity as finding a theory incorporating perturbative quantum gravity in some way or another which is UV-better, i.e. more predictive at high energies than perturbative quantum gravity; the problem is not necessarily to arrive at an `UV-complete' theory, i.e. a theory that stays formally predictive up to all energies.

The theory on the AdS side---superstring theory---is now one of the main contending approaches for such an UV-better integration between GR and QFT\footnote{Just as loop quantum gravity for instance must aim at being (at least) a UV-better theory.} but standardly only admits a perturbative formulation. The AdS/CFT correspondence, however, promises a non-perturbative, and thus UV-better formulation in terms of the CFT side: the AdS/CFT correspondence ``is itself our most precise definition of string theory, giving an exact construction of the theory with $AdS_5 \times S^5$ boundary conditions" \citep[182]{horowitz_gauge/gravity_2009}. Thus, there is a minimal sense in which a theory of quantum gravity can feature through the AdS/CFT correspondence. This type of role of the AdS/CFT correspondence in the development of string theory is what \citet{de_haro_heuristic_2018} denotes the \textit{theoretical function} of dualities. On this view, the dualities are assumed to be exact, but they can still be employed in ``extracting the content of a theory `that is somehow already there', even if only implicitly, using a set of rules" \citep[10]{de_haro_heuristic_2018}. The isomorphism entailed by the conjecture that the AdS/CFT correspondence is an exact duality can be exploited to better our understanding of one side of the duality using the other; an effect that is only further amplified by the AdS/CFT correspondence being a weak/strong duality \citep[18]{de_haro_heuristic_2018}.

However, the scope of the AdS/CFT correspondence seems to extend even further since it relates a theory of gravity to a gauge theory.\footnote{In particular cases this is a relation between semi-classical gravity and $\mathcal{N}=4$ SYM, though the more general formulation of the gravity side is as type IIB string theory.} This appears to open new avenues from the perspective of quantum gravity for understanding the origin of gravitational degrees of freedom in non-gravitational quantum degrees of freedom; as already suggested by the remarks of \citet[34-35]{natsuume_ads/cft_2015} and \citet[3]{rangamani_holographic_2017}. One might therefore hope that the AdS/CFT correspondence suggests a \textit{more general} prescription of how gravity emerges from non-gravitational quantum degrees of freedom. \citet[2]{hubeny_ads/cft_2015} for instance writes in a recent review of the AdS/CFT correspondence: 
\begin{quote}
    This new type of holographic duality not only provided a more complete formulation of the theory, but also profoundly altered our view of the nature of spacetime: the gravitational degrees of freedom emerge as effective classical fields from highly quantum gauge theory degrees of freedom. This harks back to earlier expectations motivated by black hole thermodynamics, that spacetime arises as a coarse-grained effective description of some underlying microscopic theory, but with a new twist: the relevant description is lower-dimensional.
\end{quote}
The claim we want to defend in this section is that the AdS/CFT correspondence does not \textit{without alteration} provide such a new prescription of quantum origin of gravity. The AdS/CFT correspondence thus only offers itself as a theory of quantum gravity in the minimal sense of one side of the duality being a theory of quantum gravity; for the variant of AdS/CFT correspondence considered this is a type IIB string theory on asymptotically $AdS_5 \cross S^5$, i.e. as potentially providing a ``more complete formulation" of this type of string theory.

\subsection{AdS/CFT correspondence as alternative representation}

\label{section:31}

The problem lies in that the AdS/CFT correspondence involving a duality, albeit with additional interpretive commitments, entails an isomorphism between the AdS side and the CFT side such that they are different representations of the same bare theory. That this and other isomorphisms obtain is a mathematical fact and not a possible fact about the way the world is. When Alice counts the swans in the lake using exponentials and multiplication, she it not relying on any deep insights about swans, lakes, or the world in general, but rather on an insight about the relation between $(\mathbb{R},+)$ and $(\mathbb{R}^+ , \cdot )$. But for the same reason, it cannot be a deep insight about the way the world is that Alice can count swans like this; particular expressions about the relation between Alice's and ordinary peoples' ways of counting---such as $4 \text{ swans} \stackrel{dual}{=} \exp{4} \text{ swans}$---are (conjectured) analytic truths.\footnote{Assuming that relations among mathematical representations are analytic. We will leave the precise terminological issue to the philosopher of mathematics.} This also applies to the duality relations coming out the AdS/CFT correspondence: They are not candidates for deep insights about the way the world is; though they might be very profound insights about the relations among different representations of reality.

In particular, in a world where the outcome of any experiment can be described by the AdS side there is also always (in principle) an alternative account of these experiments in terms of the CFT side and \textit{vice versa}, assuming their conjectured duality; though this alternative representation might not be known.\footnote{Being a weak/strong coupling, only one side of the duality is actually tractable in most cases. This is only further amplified by the absense of a non-perturbative formulation of the AdS side.} 
No additional or external assumptions about the realisation of the AdS/CFT correspondence go into this, since the duality aspect of their relation is a purely formal relation. 
As such, there is not an additional fact about the realisation of the duality aspect of the AdS/CFT correspondence that might or might not obtain in the world beyond the question whether the CFT or AdS description obtains (as was the case for the isomorphism between $(\mathbb{R},+)$ and $(\mathbb{R}^+ , \cdot )$). The same goes in general for any gauge/gravity duality.

A similar point is made by Dean \citet[317]{rickles_ads/cft_2013}, who observes 
\begin{quote}
    that the duality relation is formally symmetric, so it would apparently make just as much sense to say that the gauge theory emerges from gravitational theory as the other way around.\footnote{Similar observations are made by \citet{teh_holography_2013} and \citet{de_haro_dualities_2017}.}
\end{quote} 
But rather than expressing this is terms of the symmetry between the two sides of the duality, we will emphasise that the absence of direct import from the AdS/CFT correspondence for the quantum origin of gravity derives from its inter-representational nature. This becomes clear through contrasting an (inter-representational) area law from the AdS/CFT correspondence---the Ryu-Takayanagi formula presented in the previous section---to the Bekenstein-Hawking formula \citep{bekenstein_black_1973} (which is an intra-theoretical relation). 
The Bekenstein-Hawking formula relates the horizon area, $A_{BH}$, of a stationary black hole to its entropy, $S_{BH}$:
\begin{equation}
    S_{BH} = \frac{A_{BH}}{4 G_N \hbar}
\end{equation}
\noindent
Comparing to the Ryu-Takayangi formula (see equation \ref{eq:RyuTakayanagi}), the similarity is striking. Both relate entropies to areas in terms of the same proportionality constant. But while the Bekenstein-Hawking formula applies only to certain horizon structures (in particular that of black holes), the Ryu-Takayanagi formula is a generic relation between areas on the AdS side and (entanglement) entropies on the CFT side. 
Still, it might be alluring to conceive of the Ryu-Takayanagi formula as a generalisation of the Bekenstein-Hawking formula. The Ryu-Takayanagi formula might even be interpreted to signify that the Bekenstein bound on entropy is always saturated in the AdS/CFT correspondence and perhaps even in quantum gravity in general. But such conclusions have to be resisted. In the following we explore further where the analogy between the Bekenstein-Hawking formula and the Ryu-Takayanagi formula fails---a comparison that is new to the philosophical literature---since this nicely illustrates the limitations of the AdS/CFT correspondence in general.

Even though the status of entropy in the Bekenstein-Hawking formula is arguably conceptually problematic,\footnote{Cf. for instance \begin{quote}
    it seems far from clear as to whether we should think of these degrees of freedom as residing outside of the black hole (e.g., in the thermal atmosphere), on the horizon (e.g., in Chern-Simons states), or inside the black hole (e.g., in degrees of freedom associated with what classically corresponds to the singularity deep within the black hole) \citep[31]{wald_thermodynamics_2001}.
\end{quote} It is beyond the scope of this paper to go into the details of these different accounts of black hole entropy. We refer the reader to \citet{wald_thermodynamics_2001} and \citet{carlip_black_2014}; and for a more philosophical discussion on the status of black hole thermodynamics as thermodynamics \textit{proper} to \citet{DoughertyCallender} and \citet{WallaceThermodynamics}.} all accounts locate the degrees of freedom associated with the entropy in the same spacetime as that of the black hole horizon.
The Bekenstein-Hawking formula, in other words, states a relation among quantities in the same representation in which the black hole finds itself, and is thus an intra-theoretical relation. As such, it signifies an intriguing discovery in the context of black holes about the relation between a thermodynamic property related to the number of degrees of freedom, entropy, and spacetime, in the form of the horizon area.

This is in stark contrast to the Ryu-Takayanagi formula. Here the entropy is associated with some CFT degrees of freedom.\footnote{It happens to be entanglement entropy but this is not the important contrast with the Bekenstein-Hawking formula.} But in the CFT representation where these degrees of freedom live, the co-dimension two-surface, $\tilde{B}$, whose area they relate to is nowhere to be found. Instead, this is a surface in the alternative representation of the same bare theory in terms of the AdS side. Whereas the Bekenstein-Hawking formula is the surprising discovery of a relation between very distinct quantities in the same representation (an intra-theoretical relation), the Ryu-Takayanagi formula relates two different representations of the same bare theory to one another (an inter-representational relations). 
The Ryu-Takayanagi formula is \textit{not} a relation between the entropy of a volume and the surface area of this volume but rather a relation between the entropy in one representation and a surface in another. It is the equivalent of discovering that you can count and add both like your 9-year old alter ego and like mathematically gifted Alice---using an exponential sequence, and multiplication instead! This might be (very) useful, but it is not in itself a profound insight about reality. And this generalises to all other relations coming out of the AdS/CFT correspondence; duality relations cannot directly attest to the origin of back hole degrees of freedom---as alluded to by \citet[34-35]{natsuume_ads/cft_2015} above---nor the origin of any other gravitational degrees of freedom. They simply attest to the discovery that the very same bare theory---and the same physics under the interpretation as a correspondence---has two alternative representations, and how these representations relate to each other.

To drive our point home: 
Any relation coming out of the AdS/CFT correspondence is---modulo significant differences in complexity and transparency---analogous to the truism `all bachelors are unmarried men'. 
The relation between bachelors and unmarried men holds irrespective of the way the world is, but this also means that it is not an insight about reality: while this relation is helpful to anyone not knowing what a bachelor is it cannot directly do more than lifting such kind of ignorance.\footnote{Having only a perturbative formulation of the AdS side is analogous to not knowing well what a bachelor is; each of these cases of ignorance can be shed light on by exploring the arguably easier-to-deal-with CFT side and conceptualization in terms of unmarried men, respectively.} 
If we established in some way or another that there are bachelors, then there is no additional question as to whether all these bachelors are unmarried men. Similarly, it is nonsense to ask whether the existence of unmarried men causes the existence of bachelors or whether bachelors emerge as a consequence of unmarried men. Unmarried men are neither more nor less real or fundamental than bachelors. It is a relation among representations like an exact duality. 

So, while the AdS/CFT correspondence can serve to improve our understanding of either side of the duality using the other (for purposes of quantum gravity most likely using the CFT side to learn about the string theory on the AdS side), nothing else can be achieved directly by the AdS/CFT correspondence qua duality; particularly not concerning the \textit{emergence} of gravity from quantum degrees of freedom in any other theory.\footnote{It is still a profound insight but one about the surprising relation between two theoretical frameworks that were hitherto conceived to be distinct.} Rather, it must (and, as we will show, does) result from highly non-trivial heuristics if the AdS/CFT correspondence---as \citet[2]{hubeny_ads/cft_2015} claims---shows how ``spacetime arises" from a ``lower dimensional" quantum theory. 

\section{From inter-representational to intra-theoretical relations: Two case studies}

To activate the discoveries of the AdS/CFT correspondence for general advances in the search for the quantum origin of gravity, one would have to seek means for embedding its inter-representational relations into one and the same theory, that is to promote inter-representational relations into intra-theoretical ones. 
As alluded to above, this goes beyond rendering the duality in-exact; the methodology of holography without holography, developed below, does this by keeping aspects of the formal framework of the AdS/CFT correspondence while leaving holography, and thus the inter-representational relations, behind. Holography without holography follows the spirit of the heuristic function of dualities as explicated by \cite{de_haro_heuristic_2018} where dualities are used to build new theories.

The methodology of holography without holography amounts to a particularization of this heuristic function of dualities where the symmetric duality relations of the AdS/CFT correspondence serve as guiding principles in the search for the relation between underlying quantum degrees of freedom and gravity \textit{without}, however, realising the AdS/CFT correspondence. We demonstrate in this section how the works of Jacobson and Verlinde respectively manage to do this and thereby implement holography without holography, though only implicitly so. These case studies then nicely illustrate the point already made above that this heuristic use of the AdS/CFT correspondence cannot be achieved by simply stipulating that the relations of the AdS/CFT correspondence are not exact. As, for instance, Jacobson's implementation of the Ryu-Takayanagi formula demonstrates, the problem is not that the relation---and the duality as a whole---is exact (and thus symmetric), but rather how to embed the entropy and the area in the same theory. The methodology of holography without holography provides for a procedure towards this end.

\subsection{Jacobson 2016: Entanglement Equilibrium and the Einstein field equation} 

Ted \citet{jacobson_entanglement_2016} promises that
\begin{quote}
The present work combines the local spacetime setting of the equation of state approach [as in \citep{jacobson_thermodynamics_1995}], with the statistical, compact-region setting of the holographic analysis [as in \citet{faulkner_gravitation_2014}], but it proceeds directly in spacetime, making no use of holography \citep[2]{jacobson_entanglement_2016}.
\end{quote}
The mentioned ``holographic analysis" is the derivation of linearised Einstein field equations from the first law of entanglement entropy in the context of the AdS/CFT correspondence by van Raamsdonk and collaborators  \citep{van_raamsdonk_building_2010,lashkari_gravitational_2014,faulkner_gravitation_2014}). The mentioned ``equation of state approach" is Jacobson's \citeyear{jacobson_thermodynamics_1995} own thermodynamic take on gravity in which the Einstein field equations are derived from a thermodynamic-like equation of state. Motivated by these two programs, Jacobson manages to derive the Einstein field equations from an entanglement equilibrium condition. In the following, we will first depict the derivation of van Raamsdonk and collaborators in the holographic setting, and then show how Jacobson manages to invoke the spirit of this derivation in a completely non-holographic setting.

\subsubsection{The holographic version}\label{Holographic}
In this section, we review the equivalence of constraints on the entanglement entropy on the CFT side to the validity of the linearised Einstein field equations on the AdS side as considered by \citet{lashkari_gravitational_2014,faulkner_gravitation_2014,swingle_universality_2014}. We in particular emphasise those aspects of the derivation of this equality that are paralleled by Jacobson's non-holographic derivation of the field equations.

In any QFT, entanglement entropy features in a first-law-like expression (`first law of entanglement entropy'):
\begin{equation}\label{dSB=dtr}
\delta S_B  = \delta \expval{H_B} 
\end{equation}
where $B$ is some (sub)region of the domain of the QFT, $\delta S_B$ is the first order variation of the entanglement entropy of this region \textit{away from the vacuum state of the QFT}, and $\delta \expval{H_B}$ is the first order variation of the expectation value of the modular Hamiltonian $H_B \equiv - \log(\rho_B^{Vac})$ ($\rho_B^{Vac}$ denotes the vacuum state linked to region $B$). The modular Hamiltonian is the Hamiltonian with respect to which the state $\rho_B^{Vac}$ can be expressed as a thermal state, that is $\rho_B^{Vac}=\exp(-H_B)$. Though the first law of entanglement entropy looks like the Clausius relation for thermal systems (`$T dS = dE$'), it is a general result for any QFT---in particular, it does not presuppose an equilibrium(-like state).

By the Ryu-Takayanagi formula, $\delta S_B$ can be related to the variation of the area of the co-dimension two surface $\tilde{B}$ in the dual representation (see figure \ref{cylinder}) such that
\begin{equation}\label{AB=HB}
    \frac{ \delta A(\tilde{B})}{4 G_N \hbar} \stackrel{dual}{=} \delta \expval{H_B}.
\end{equation}
This relation, however, is in general not of much use since the modular Hamiltonian, $H_B$, is generally not a local operator.\footnote{A local operator is an operator that can be expressed in terms of an integral of the quantum fields and their derivatives.} But if $B$ is a ball-shaped region\footnote{More precisely, the ball-shaped region is given by what is usually called a (spacelike) geodesic ball of radius $l$ with center $p$. See for instance \citet[appendix A]{jacobson_entanglement_2016}.} in Minkowski spacetime, then there exists a conformal mapping between the state $\rho_B^{Vac}$ and the corresponding state $\rho_{\eta}$ as seen by a constantly accelerating observer (called the `Rindler observer') \citep{casini_towards_2011}. In a conformal field theory, the operators are invariant under conformal transformation, and we may therefore use this mapping to express the modular Hamiltonian in terms of the boost Hamiltonian\footnote{Also known as the Rindler Hamiltonian.} associated with the accelerating observer. In doing so, we find a local expression for $\expval{H_B}$ as an integral of the energy-momentum tensor over the ball-shaped region $B$,\footnote{For completeness the full expression is: \begin{equation}\label{HB=Ttt}
\expval{H_B} = 2 \pi \int_B d^{D-1} x \frac{R^2-(\vec{x}-\vec{x}_0)^2}{2R} \expval{T^{tt}(x)} \equiv E^{Hyp}_B
\end{equation} 
where $T^{tt}(x)$ is the energy density of the CFT and $R$ is the radius of the ball shaped region, $B$.} which we may identify as some hyperbolic energy, $E^{Hyp}_B$, of the quantum state on $B$. But to re-emphasise, this expression of the modular Hamiltonian as a local hyperbolic energy only holds in a conformal field theory.\footnote{For later purposes, observe that for an infinitesimally small ball-shaped region $\delta \expval{T^{tt}(x)}$ may be approximated by a constant throughout the ball which allows the evaluation of the integral (\ref{HB=Ttt}) for the first order variation:
\begin{equation}\label{Hinf=Ttt}
\delta \expval{H_B^{Inf}} = 2 \pi \frac{\Omega_{D-2} R^D}{D^2-1} \delta \expval{T^{tt}(x_0)}
\end{equation} 
where $H_B^{Inf}$ is the modular Hamiltonian of the infinitesimal ball and $\Omega_{D-2}$ is the surface area of a sphere of dimension $D-2$.} Inserting this into eq. (\ref{AB=HB}), we obtain a relation between an area variation on the AdS side and an energy variation on the CFT side:
\begin{equation}\label{AB=EB}
    \frac{ \delta A(\tilde{B})}{4 G_N \hbar} \stackrel{dual}{=} \delta E^{Hyp}_B.
\end{equation}

It is a general result of the AdS/CFT correspondence that energies of the CFT state correspond to energies of the dual spacetime \citep{balasubramanian_stress_1999}.\footnote{Here the energy of a spacetime is interpreted as some quasi-local energy momentum tensor, since any local operator depending only on the metric and its first order derivatives must vanish in a generally covariant theory.} We can therefore obtain an interpretation of the CFT energy $E^{Hyp}_B$ on the AdS side, and thus relate the AdS area variation to an AdS energy variation. In a final step, this can then be shown (as done by \citet{lashkari_gravitational_2014} and \citet{faulkner_gravitation_2014}) to be equivalent to the linearised vacuum Einstein field equations with negative cosmological constant (that are solved by AdS spacetime). Moreover, adding first order corrections to Ryu-Takanagi formula due to bulk entanglement over $\tilde{B}$ will give the linearised, semi-classical Einstein field equations with negative cosmological constant; it is speculated that the inclusion of all orders in $\delta S$ and $\delta E^{Hyp}_B$ will give the full semi-classical Einstein field equations \citep{swingle_universality_2014}.

In sum, starting from the formal result $\delta S_B  = \delta \expval{H_B}$, the left hand side, $\delta S_B$, receives its holographic interpretation as an area on the AdS side using the Ryu-Takayanagi formula. The right hand side, $\expval{H_B}$, is first recognised as a local hyperbolic energy using a conformal mapping and then translated to the AdS using the general relation between CFT and AdS energy. The resulting local constraint relating the area variation to energy (both on the AdS side) is then shown to be equivalent to the linearised Einstein field equations.

\subsubsection{The non-holographic version}
The outset for Jacobson's derivation of Einstein's field equation is a quantum state $\ket{\Psi}$ defined on a maximally symmetric spacetime,\footnote{The maximally symmetric spacetimes are Minkowski spacetime and (Anti-)de-Sitter spacetime.} $(\mathcal{M}, g_{\mu \nu})$. One then considers the domain of dependence---the causal diamond---of a ball shaped region, ${\mathbb{B}}$, (`geodesic ball') within this spacetime.\footnote{Whereas the holographic version considered a ball-shaped region on the boundary, denoted $B$, this non-holographic version considers a ball-shaped region in the bulk, which we denote ${\mathbb{B}}$ to avoid confusion.} The quantum degrees of freedom within the ball will generally be entangled with the degrees of freedom outside the ball. Thus, the ball is associated with a non-zero entanglement entropy, $S_{\mathbb{B}}^{tot}$, i.e. an entanglement across the boundary surface of ${\mathbb{B}}$, $\partial {\mathbb{B}}$. 

In the holographic version, the first step of the derivation involves finding a relation between an energy and an entanglement entropy (component) that in turn can be related to the area of a co-dimension two surface of the bulk spacetime. The first step in Jacobson's derivation lies in establishing an analogous relation while however sticking to the theory defined in $(\mathcal{M}, g_{\mu \nu})$, that is without using a holographic relation. This entails that the featured co-dimension two-surface cannot go into \textit{some kind of bulk} (relative to which the theory $(M, g)$ `sits' on the boundary) but must be identified with a co-dimension two-surface in $\mathcal{M}$ itself. And similarly, the entanglement entropy cannot be associated with that of a different representation (living on the boundary) of the system. In contrast to the holographic version, this setting is intra-theoretical rather than inter-representational; the inter-representational derivation is only used as a guiding principle in the generation of an analogous intra-theoretical derivation that does not assume holography. This is the methodology of holography without holography and we shall return to this role of it as a guiding principle in section 5.

In the non-holographic setting, the most obvious candidate for a co-dimension two-surface is the boundary of ${\mathbb{B}}$, $\partial {\mathbb{B}}$. Indeed, it is a generic result of quantum many body systems that the entanglement over the boundary of a subsystem (such as $\partial {\mathbb{B}}$) scales (to leading order) with the area of the boundary (in this case, the area of $\partial {\mathbb{B}}$) \citep{eisert_colloquium:_2010}. This leading order behavior is primarily due to vacuum fluctuations in the near vicinity of $\partial {\mathbb{B}}$ governed by the UV physics and not the state of the subsystem. We can therefore conceive of this as a UV contribution to the total entanglement entropy that scales with the area of the boundary and is independent (to leading order) of the state of the subsystem: $S_{{\mathbb{B}}}^{UV} = \eta A(\partial {\mathbb{B}})$, where $\eta$ is some proportionality constant. Any other contributions to the total entanglement across $\partial {\mathbb{B}}$ is due to IR (long range) physics and is determined by the state of the subsystem. This is, in other words, entanglement across $\partial {\mathbb{B}}$ that is not originating in the entanglement between degrees of freedom immediately in the vicinity of $\partial {\mathbb{B}}$. Thus, the total entanglement entropy relative to the vacuum between the quantum subsystem on $\mathbb{B}$ and the rest of the system can be expressed as:
\begin{equation}\label{SBtot}
    S_{\mathbb{B}}^{tot} = S_{\mathbb{B}}^{UV} + S_{\mathbb{B}}^{IR} = \eta A(\partial {\mathbb{B}}) + S_{\mathbb{B}}^{IR}.
\end{equation}
Consider a simultaneous variation of the metric and the quantum state $(\delta g_{\mu \nu}, \delta \ket{\Psi})$ away from the vacuum state.\footnote{Notice that we do two simultaneous variations here, whereas in holographic version we only varied the CFT state which entailed a consequent variation of the AdS spacetime metric.} If $\eta$ is a constant (as is assumed), and if it holds that $S_{\mathbb{B}}^{UV} = \eta A(\partial {\mathbb{B}})$ then the variation of $S_{\mathbb{B}}^{UV}$ only depends on $\delta g_{\mu \nu}$ since $A(\partial {\mathbb{B}})$ is independent of $\ket{\Psi}$ to leading order.\footnote{To leading order, we can disregard backreaction from the matter fields on the metric.} The variation of $S_{\mathbb{B}}^{IR}$, on the other hand, depends only on $\delta \ket{\Psi}$ since it is, to leading order, independent of the exact location of the boundary. We therefore get:
\begin{equation}\label{dSBtot}
    \delta S_{\mathbb{B}}^{tot} = \eta \delta A(\partial {\mathbb{B}}) + \delta S_{\mathbb{B}}^{IR}.
\end{equation} 

As stated above, it is a general result of any QFT that the variation of the entanglement entropy can be related to the variation of the expectation value of the modular Hamiltonian when the variation is taken with respect to the vacuum state. However, since $S_{\mathbb{B}}^{UV}$ is independent of a variation of the state, it follows that only the variation of the IR component of the entanglement entropy contributes to the variation of the modular Hamiltonian, i.e. $\delta S_{\mathbb{B}}^{IR} = \expval{H_{\mathbb{B}}}$. We therefore have:
\begin{equation}
    \delta S_{\mathbb{B}}^{tot} = \eta \delta A(\partial {\mathbb{B}}) + \delta \expval{H_{\mathbb{B}}}.
\end{equation} 

Finally, Jacobson assumes what he calls the maximal vacuum entanglement hypothesis\footnote{For fixed volume, a thermodynamic system at equilibrium has minimal free energy, that is $\partial F = \partial E - T \partial S = 0$. Since the energy of a maximally symmetric spacetime is zero, minimisation of free energy corresponds to maximisation of entropy.}: 
\begin{quote}
    When the geometry and quantum fields are simultaneously varied from maximal symmetry, the entanglement entropy in a small geodesic ball is maximal at fixed volume \citep[1]{jacobson_entanglement_2016}.
\end{quote}
The first order variation of entanglement entropy away from the vacuum state vanishes for fixed volume, i.e. $\delta S_{\mathbb{B}}^{tot}=0$ when the variation of the geometry and quantum state is such that the volume of the subsystem is unchanged. From this, we obtain: $ 0 = \eta \delta A(\partial {\mathbb{B}}) + \delta \expval{H_{\mathbb{B}}}$ closely resembling eq. (\ref{AB=HB}) from the holographic version.\footnote{Note that the variation of area is at fixed volume as opposed to fixed radius. Comparing to the holographic analysis of Van Raamsdonk and others (see the previous section), this is a special case where the state of the system is assumed to be such that the \textit{energy} of the ball vanishes (as the variation happens around a maximally symmetric spacetime for which in fact $E=0$). However, the derivation still runs in parallel.} However, this time the area variation relates to a modular Hamiltonian in the same representation: this is an intra-theoretical rather than inter-representational relation.

The issue, however, is the same: $H_{\mathbb{B}}$ is generally not a local operator. Had this been a conformal field theory, we could directly proceed by using the conformal mapping of this state from a ball shaped region to the state as seen from the Rindler observer system and express $\expval{H_{\mathbb{B}}}$ in terms of the Rindler Hamiltonian. As in the holographic version above, we could thereby find a local expression for $\expval{H_{\mathbb{B}}}$ as an integral of the energy-momentum tensor over the ball-shaped region $\mathbb{B}$, which we may identify as some hyperbolic energy, $E^{Hyp}_{\mathbb{B}}$, of the quantum state on ${\mathbb{B}}$.
Although we cannot generally expect to have conformal invariance, \citet[4]{jacobson_entanglement_2016} speculates that the matter considered here is suitably described by a theory with a UV fixed point, and, furthermore, that the energy regimes of interest are sufficiently close to that of the fixed point such that it can in fact be treated as approximately conformal for small length scales compared to the characteristic length scale of the quantum field theory in question.\footnote{See \citet{cao_bulk_2018} for another non-holographic derivation of the Einstein field equations that is inspired by the holographic derivation of section \ref{Holographic} but which trades the requirement of a UV-fixed point for other assumptions.} Jacobson therefore assumes that the modular Hamiltonian can be expressed in terms of a hyperbolic energy of ${\mathbb{B}}$, $E^{Hyp}_{\mathbb{B}}$, with some spacetime scalar correction, $X$, that must be assumed to be small.\footnote{As it turns out, $X$ is related to the curvature scale of the spacetime such that the derivation as a whole is only valid if $X$ vanishes or is small everywhere compared to $E^{Hyp}_{\mathbb{B}}$ which signifies the importance of the assumption that the degrees of freedom are close to conformal.} The IR entanglement entropy can therefore be expressed as: $\delta S_{{\mathbb{B}}}^{IR} = \delta \expval{H_{\mathbb{B}}} = \delta E^{Hyp}_{\mathbb{B}} + \delta X$.\footnote{For completeness, we have:
\begin{equation}\label{HIR=Ttt}
    \delta S_{\mathbb{B}}^{IR} = \delta \expval{H_{\mathbb{B}}} = \frac{2 \pi}{\hbar} \frac{\Omega_{D-2} R^D}{D^2-1} \left( \delta \expval{T_{tt}(x_0)} + \delta X \right) \equiv E^{Hyp}_{\mathbb{B}}
\end{equation}
 Compare with eq. (\ref{Hinf=Ttt}).} Thus, assuming that the fundamental quantum degrees of freedom are sufficiently similar to CFT degrees of freedom, we can obtain a relation similar up to a sign and small scalar to that obtained in the successful derivations of Einstein field equations in the AdS/CFT correspondence: 
\begin{equation}
    0 = \eta \delta A(\partial {\mathbb{B}}) + \delta E^{Hyp}_{\mathbb{B}} + \delta X.
\end{equation} 

From this constraint, Jacobson then derives the full Einstein field equations by a procedure similar to that of \citet{faulkner_gravitation_2014}. Jacobson can derive the full Einstein equation and not only the linearised equation due to differences in the expressions of the area variation.\footnote{In essence, Jacobson can help himself to more assumptions since he is not constrained by the framework of the AdS/CFT correspondence.}

In building on the Ryu-Takayanagi formula and thus holography, the derivation of van Raamsdonk and collaborators merely show that the Einstein equations on the AdS side are equivalent to an entanglement constraint on the CFT side. Despite its promise to relate gravity and entanglement, the derivation is simply a formal result within the AdS/CFT framework; it signifies how to translate the Einstein field equations when moving from the gravity representation on the AdS side to the non-gravitational representation on the CFT side. Jacobson reproduces---with some deviations---this derivation, but replaces any components coming from the AdS/CFT correspondence with plausible non-holographic assumptions. First, the inter-representational Ryu-Takayanagi formula---needed to have a relation to spacetime in the first place---is replaced by a conjectured leading order proportionality between the area and entanglement entropy of the same ball; an intra-theoretical relation. Second, since it is only part of the entanglement that is related to area, Jacobson has to help himself to the additional assumption that the total entanglement entropy vanishes, but together these two assumption reproduce, up to a sign, eq. (\ref{AB=HB}) from the holographic version in a non-holographic setting. Third, assuming that the microscopic degrees of freedom are close to conformal---a property that followed directly in the holographic version, the CFT being conformally invariant---Jacobson reproduces, up to a sign and a small scalar, eq. (\ref{AB=EB}) from which the Einstein field equations can be derived. Whereas the holographic version shows that the Einstein field equations on the AdS side are equivalent to an entanglement constraint on the CFT side, an inter-representational relation, Jacobson promotes this into a relation between the microscopic quantum degrees of freedom and an emergent leading order gravitational dynamics that accords with GR. More precisely, the (apparent) gravitational dynamics is collective behavior resulting from entanglement thermodynamics in the form of the hypothesis that the vacuum entanglement is maximal ($\delta S_{\mathbb{B}}^{tot}=0$), i.e. at an equilibrium. Indeed, this result that a small ball---a small local system---should be at an entanglement equilibrium after some time is supported by quantum thermalisation in generic many-body systems \citep{kaufman_quantum_2016}. The duality between the Einstein field equations and an entanglement constraint derived in the holographic version informs a thermodynamic coarse-grained relation in the non-holographic setting where entanglement equilibrium, assumed UV-completion, and leading order area scaling of the entanglement entropy replace the inter-representational relations of the holographic version.\footnote{
As such, Jacobson's derivation might be conceived of as realizing the anticipated role of the AdS/CFT correspondence as a correspondence principle à la Bohr that \citet{van_dongen_emergence_2019} trace back to the inception of AdS/CFT by \citet{maldacena_large-n_1999}. Similarities and dissimilarities between this case and that of Bohr and his contemporaries might be an interesting venue for further research.
}

\renewcommand{\arraystretch}{2}

\begin{table}[]
    \centering
    \begin{tabular}{p{4cm}|p{5cm}|p{5cm}}
         & Van Raamsdonk (with holography) & Jacobson (without holography)\\
         \hline
         Entanglement entropy linked to Hilbert space of & quantum system defined on region $B$ of the boundary, $\partial M_{\Psi}$ & quantum system defined on a region $\mathbb{B}$ in the bulk\\   

         Balance equation & $\delta S_B  = \delta \expval{H_B}$ & $\delta S_{\mathbb{B}}^{tot} = \delta S_{\mathbb{B}}^{UV}   + \delta \expval{H_{\mathbb{B}}} = 0$\\
         Area law & $S_B \stackrel{dual}{=} \frac{ A(\tilde{B})}{4 G_N \hbar} $  & $S_{{\mathbb{B}}}^{UV} = \eta A(\partial {\mathbb{B}})$
    \end{tabular}
    \caption{Comparison of basic elements in the derivation of the Einstein field equations in the holographic (van Raamsdonk) and non-holographic (Jacobson) version.}
    \label{JvRtable}
\end{table}

\subsection{Verlinde 2017: Via tensor networks to entropy in the bulk}

Erik \citet{verlinde_emergent_2017} also explicitly draws on the relations between spacetime and entanglement coming out of the AdS/CFT correspondence. His ambition is, in accordance with the methodology of holography without holography, to let these relations inspire prospective insights about the emergence of spacetime and gravity with the goal to ``apply them to a universe closer to our own, namely de Sitter space" \citep[3]{verlinde_emergent_2017}.\footnote{Verlinde does at places speak as if AdS spacetime and gravity are emergent from the CFT degrees of freedom and thus disregards (or overlooks) the inter-representational character of the AdS/CFT correspondence. However, the reconstruction of his argument below avoids this misunderstanding.}

\begin{figure}

	\begin{center}
		\includegraphics[scale=0.4]{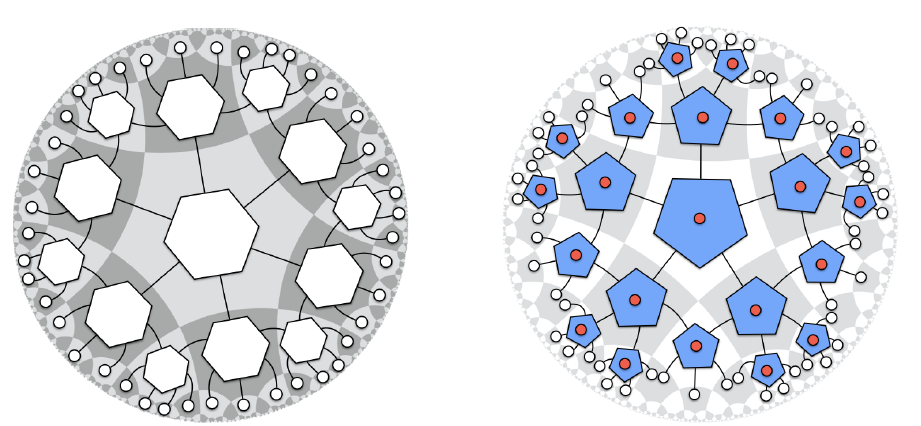}
		
	\end{center}
		
		\caption{\label{TensorNetworkAdSCFT} Illustration of a (a) holographic state (left), and a (b) holographic code (right) --- the tensor network analogue for the AdS/CFT correspondence. (This is figure 4 as taken from \cite{Pastawski}.) 
		} 

\end{figure}

Verlinde's argument builds on a remarkable relation between the AdS/CFT correspondence and the multi-scale entanglement renormalization ansatz (MERA) known from quantum many-body systems.\footnote{The presentation here largely follows \citet[ch. 14]{rangamani_holographic_2017}.} In its original context, MERA can be regarded as a coarse-graining schema: If we consider a chain of $n$ spins, MERA involves joining $p$ neighbouring spins into one block such that the number of spins at this first level of coarse-graining, (call it $u=1$), is $n/p$. Notably, this scheme can be applied iteratively until only one block remains such that after $u=2,3,4,...$ steps of coarse-graining, the number of spins is $n/p^u$. An associated mapping \textit{from} the coarse-grained \textit{to} the fine-grained Hilbert space is to be described by an isometric tensor, i.e. a tensor that preserves the norm of the quantum states. A pseudo-inverse to this isometric tensor then serves as a `coarse-graining tensor' that joins $p$ spins into one.\footnote{This is basically the block spin renormalisation group as championed by \cite{Kadanoff}.} For each additional level of coarse-graining, $u=2,3,4,...$, another coarse-graining tensor is required. Together, these coarse-graining tensors compose a map from a fine-grained state to a respectively more coarse-grained level $u=2,3,4,...$. (Or seen conversely again, the corresponding isometric tensors compose a map from a more-coarse-grained level to a more fine-grained state.) 
Such a map can be depicted graphically as a network where each isometric/coarse-graining tensor (the nodes) takes one input (referred to as the `incoming legs') and delivers $p$ outputs (called the `outgoing legs') each of which in turn serves as input for other tensors (see figure \ref{TensorNetworkAdSCFT}).\footnote{More generally, a tensor $T$ with indices $\alpha \beta ... a b ...$ can be used to express quantum states such as $\ket{\psi} = \sum_{\alpha, \beta, ..., a, b, ... =1}^D T^{\alpha \beta ... a b ...} \ket{\alpha} \ket{\beta} ... \ket{a} \ket{b} ...$ --- but also to depict maps from one Hilbert space $\mathcal{H}_A$ into another Hilbert space $\mathcal{H}_B$ with $\text{dim}(\mathcal{H}_A) \leq \text{dim}(\mathcal{H}_B)$ such as $\ket{\alpha} \ket{\beta} ... \mapsto \sum_{a, b, ... =1}^D T^{\alpha \beta ... a b ...} \ket{a} \ket{b} ...$. It is the latter use we have in mind here. In a graphical representation of such a tensor-induced map, one would call the indices $\alpha, \beta, ...$ incoming legs, and the indices $a, b, ...$ outgoing legs (see \cite{Pastawski}). Tensor \textit{networks} are then graph structures where each node is associated with such a tensor (and only one tensor), and each edge between nodes with a summation of a joint index of two tensors. Note that norm-preserving tensors can have at most as many incoming legs as outgoing legs; a norm-preserving tensor can also be (non-uniquely) pseudo-inverted and then thought of as a coarse-graining operation --- as we are doing here partly. An inverted norm-preserving tensor has at least as many incoming legs as outgoing legs.} The legs are in other words the edges connecting the nodes.\footnote{In the formalism, each edge corresponds to a summation of joint indices between the tensors connected by the edge.} In this tensor network picture, the boundary of the network corresponds to the map between the original (fine-grained) state, $u=0$, and the first coarse-grained level, $u=1$. As one moves into the bulk, one goes towards higher $u$ and towards maps between ever more coarse-grained levels.

In order for the coarse-grained states to be renormalizable, all entanglement must be removed between neighboring blocks as one moves inwards through the network and thus towards a more coarse-grained description. This information about the difference in entanglement across levels is ultimately what is carried by the tensors; the network thus encodes the entanglement structure of the original fine grained state that is \textit{removed} by the coarse-graining. More concretely: Moving outward towards the boundary of the tensor network (from the coarse-grained to the fine-grained description of the system), the tensors linked to the map from the level $u$ to $u-1$ both break up the blocks at the level $u$ into their constituents at the level of description of $u-1$ and reinstate the additional short range entanglement at $u-1$ that were `removed' to render the state at the level $u$ renormalizable.

So much for the account of MERA as a coarse-graining scheme. What is interesting from the perspective of the AdS/CFT correspondence is that the tensor network has an intriguing similarity to a discretised AdS spacetime when the original (boundary) state is a CFT state \citep{swingle_entanglement_2012}. A \textit{first} indication of this is gathered from the alluring similarity between the depiction of a tensor network of figure \ref{TensorNetworkAdSCFT}, and a graphical depiction of hyperbolic 2-space geometry (that constant times slices of $AdS_3$ spacetimes belong to) on a flat projection plane. Instead of being a coarse-graining scale, $u$ can be reinterpreted as the radial bulk coordinate of AdS spacetime and each tensor is conceived as encoding the local geometrical neighbouring relations. The evidence for this geometric interpretation of the MERA tensor network is still tentative (see \citet{bao_consistency_2015} for a review) but includes: the matching of the length of trajectories between AdS spacetime and the tensor network (as captured by the number of links crossed), the reproduction of a discretised version of the Ryu-Takayanagi formula, and the occurrence of error-correcting features known from the actual AdS/CFT correspondence. In summary, the geometric interpretation of the tensor network is corroborated by elements of the AdS/CFT correspondence and the similarity between the radial AdS coordinate and the coarse-graining scale $u$.

This all provides for a discretised version of the AdS/CFT correspondence. The tensor network is interpreted geometrically while it --- due to its origin as a coarse-graining procedure --- still carries the full CFT state on its boundary. In particular, any operator acting on an incoming `leg' of a tensor in the network can be represented --- `pushed through' --- to act as an equal norm operator on the outgoing leg.\footnote{See \cite{Pastawski}, section 2.} In this way, every operator can be pushed to the \textit{boundary} such that the operator acting in the bulk can be re-represented as an operator acting on the boundary. Thus, holography obtains. (As basically already said above, we can regard the tensor network from the outside towards the inside --- then local tensors of the network --- should be understood as coarse-graining operations; or we can regard the tensor network from the inside towards the outside.)

Verlinde now aims to use this tensor network version of the AdS/CFT correspondence in an exploration of de Sitter spacetime. Essentially, his proposal consists in considering what happens when one changes the tensor network so it is no longer holographic by connecting it differently, i.e. changing what tensors are joined by each edge. From the perspective of the coarse-graining scheme, this would be nonsensical: The tensor network depends on and is tailored to the original state and should be subject to change only when the state is changed. An arbitrary change in the bulk of the network would most likely entail that the approximation breaks down in the sense that the network would no longer map a coarse-grained description of some system to a fine-grained one. Once the tensor network receives its geometric interpretation, changes in the bulk of the network can be regarded---and this seems to be what Verlinde (implicitly) proposes---as changes of the geometry. The problem, however, remains that the geometric interpretation of the tensor network relies on the analogy between $u$ and the radial coordinate of the AdS spacetime and the support coming from the AdS/CFT correspondence. Neither are satisfied for arbitrary tensor networks, but rather only obtains for holographic tensor networks. 

Verlinde's proposal, therefore, is to take as the vacuum state a holographic tensor network which does admit a geometric interpretation as an AdS spacetime. From the coarse-graining perspective, the tensors of the network codified the short range entanglement that were removed between each coarse-grained level. With the geometric interpretation, the tensors no longer belong to different levels of description, but rather different depths in the bulk space. Verlinde therefore conjectures that the tensors are not transformations that reinstate entanglement, but rather entities that capture the entanglement structure underlying space(time). The network is in other words regarded as a network of entanglement (quantum information) where the links between the tensors in the network encode their mutual entanglement. This entanglement is then what gives rise to spacetime: ``spacetime geometry is viewed as representing the entanglement structure of the microscopic quantum state" \citep[3]{verlinde_emergent_2017}. Verlinde explicitly acknowledges that this idea comes from the relation between spacetime and entanglement in the AdS/CFT correspondence. His account, however, is different in an important respect (though he does not emphasize this himself): Spacetime is not dual to an entanglement structure (inter-representational), rather spacetime is conjectured to emerge from entanglement (intra-theoretical). As argued in section \ref{section:31}, this cannot simply be achieved by stipulation. Crucial for Verlinde's account therefore is the insight that the tensors can be elevated to comprise real networks of entanglement rather than transformations. Thus the dual CFT state serves to inspire, but only inspire, the quantum degrees of freedom which spacetime originates in. Re-representation in terms of a CFT state in still possible, but only when the tensor network finds itself in the configuration where it is equivalent to a MERA construction based on that CFT state; this is the holographic state that Verlinde proposes to be the ground state and which is indicated to give rise to an AdS spacetime. 

For Verlinde, therefore, the network is not dependent on the CFT state (for which the MERA is meant as a coarse-graining scheme). The holographic state is simply one configuration of the tensor network. As it turns out,\footnote{See \citet{Pastawski}, \citet{Yang}, and \citet{hayden_holographic_2016}.} the holographic state is one where each tensor is maximally entangled with its nearest neighbours; this is what allows one to push through every bulk operator all the way to the boundary. Verlinde writes: ``we take an alternative point of view by regarding all these bulk tensors as physical qubits, and interpreting the short distance entanglement imposed by the network as being due to stabilizer conditions" \citep[7]{verlinde_emergent_2017}. The short range entanglement entails that the Bekenstein-Hawking formula, the intra-theoretical proportionality between entanglement entropy and area, holds in the bulk when the stabilizer conditions are satisfied. This allows for the derivation of Einstein field equations which further corroborates the geometrical interpretation of the holographic tensor network state---the conjectured ground state---as a spacetime: ``Our interpretation of general relativity and the Einstein equations is that it describes the response of the area law entanglement of the vacuum spacetime to matter" \citep[15]{verlinde_emergent_2017}. 

Replacing some of the short range entanglement with long range entanglement in the tensor network------that is, connecting non-neighbouring tensors in the network---will have two consequences: First, the tensor network will no longer be holographic and second, the entanglement entropy will no longer satisfy an exact area law. These lay the ground for Verlinde's hope to connect with de Sitter spacetime. Since the network is no longer holographic, the geometric interpretation of the networks comes into question as suggested above. However, Verlinde proposes that long range entanglement can be treated as excitations of the holographic ground state and that they, as excited states, can thus inherit their geometric interpretation from the ground state. These excitations---in the form of changes to the entanglement of the tensor network---are thereby considered to correspond to changes of geometry. He finds further hints for this by connecting long range entanglement in the tensor network to de Sitter phenomenology and the hitherto unexplained phenomena of dark energy and dark Matter.

Generally speaking, Verlinde presents two straightforward argumentative strands for why dS space should be regarded as the ``excitation" of AdS space: one is based on entropy considerations, and the other on energy considerations; both argumentative strands build on thinking of the AdS/CFT correspondence in terms of its discrete tensor network analogue. This new conception of dS as an excitation of AdS can then, for instance, explain dark matter contributions to galaxies, which --- together with other explanatory successes --- may also be seen to corroborate this viewpoint. We will consider each of the straghtforward argumentative strands in more detail as its them which are directly related to the methodology of holography wihout holography.

One way to argue that dS is the excitation of AdS runs via entropy scaling considerations. To illustrate Verlinde's reasoning, it is best if we have the graphical representation of the tensor network analogue --- following \cite{Pastawski} --- in mind (see figure \ref{TensorNetworkAdSCFT}): consider a tiling of negatively curved AdS-spacetime in terms of hexagons and pentagons respectively. Then, across the tiling patterns, tensors with $n$ indices are placed (here, $n=6$); contraction of tensor indices gives rise to a tensor network. Read from the inside to the outside to the outside, there are always more outgoing tensor indices than incoming ones. In case of figure \ref{TensorNetworkAdSCFT} (a), each of the six indices of a tensor $x$ is contracted with that of neighbouring tensors $y$ such that mutual entanglement between each index pair is maximised (stabiliser state); schematically, this can be written as $T^{\alpha}_{\alpha} \ket{\alpha}_x \ket{\alpha}_y = \sum_{\alpha=1}^D \frac{1}{\sqrt{D}} \ket{\alpha}_x \ket{\alpha}_y$ where $x$ and $y$ denote the neighbouring tensors and $D=2, 3, ...$ for qbit/tribit/... states. In case of figure \ref{TensorNetworkAdSCFT} (b), tensor indices are contracted in the same fashion except for that one index per tensor is left open as a ``free slot". The tensor network in (b) serves as a map which transforms quantum states associated with the free slots in the bulk (red dots) into physical quantum states associated with the boundary (white dots). Thereby, the network provides a map from `red dot' inputs to `white dot' outputs. The tensor network in figure (a) is obtained from that in figure (b) by successively contracting those pairs of free bulk indices (red dots) which are neighbouring across a tiling vertex. Thereby, hexagon tilings are turned into pentagon tilings (see figure \ref{TensorNetworkAdSCFT} (b)). (Note that the graphical depictions are slightly misleading, in particular by putatively singling out the tensor in the center of the figure as special. Rather, there are graph isomorphisms for moving any specific tensor into the center without changing the local tiling structure of the graph.) Now, when long-range entanglement enters the system through thermal excitation, it dissolves the maximal entanglement structure as displayed in figure  \ref{TensorNetworkAdSCFT} (a); contractions on the level of \textit{neighbouring} tensors are then partly broken up (as the case in (b)) and replaced by contractions between \textit{non-neighbouring} tensors.\footnote{Thus, Verlinde as well as Jacobson assumes that the ground state is at an entanglement equilibrium. This might prove to be another generic feature of quantum gravity theories.} By this, the relationship between entropy and area fades away (area scaling of entropy is linked to short-range entanglement only --- as it is the case for AdS-spacetime) and volume-scaling contributions to the entropy ultimately begin to dominate --- as it is the case for dS-spacetime for large-scales.\footnote{Thanks a lot to X for clarifications on this matter.} Rather than contracting all free indices among neighbouring indices then (which gets one from (b) to (a)), the free indices should be thought of as being connected among all kind of tensors across the bulk. In particular, a re-representation of bulk states via boundary states is not possible anymore.

A second way to support the picture of dS as the excitation of AdS is based on energy-entropy considerations (see section 2.3, \citet{verlinde_emergent_2017}). First, note that the energy for a ball-shaped region of radius $r$ in AdS/dS spacetime (with \textit{AdS/dS curvature radius} $L$\footnote{The AdS and the dS metric for a static coordinate patch is given by $ds^2 = - f(r) dt^2 + \frac{dr^2}{f(r)} + r^2 d\Omega^2$ where $f(r) = 1 + \frac{r^2}{L^2}$ for AdS, and $f(r) = 1 - \frac{r^2}{L^2}$ for dS.}) is given by \[E_{(A)dS} (r) = \pm \frac{(d-1) (d-2)} {16 \pi G L^2} V(r)\] where $V(r)$ is the volume of the ball-shaped region considered. Then, using the AdS/CFT correspondence, one can partly re-express this energy expression in terms of the number of possible configurations $\mathcal{C}$ realising the corresponding CFT state. For $r=L$, the energy is provided by
\[E_{AdS} (L) = - \hbar \frac{d-2}{L} \mathcal{C} (L).\]
Similarly, one can reformulate the corresponding energy expression for dS-spacetime such that it looks---from an usual quantum mechanical perspective---like an energy state \textit{corresponding to an excitation of AdS-spacetime}:
 \[E_{dS} (L) = - \hbar \frac{d-2}{L}(\mathcal{N} (L) - \mathcal{C} (L)),\] where $\mathcal{N}(L) = 2 \mathcal{C}(L)$ tracks the additional number of configurations relative to the ground state. 
 
As a consistency check, use the energy state expression for dS qua excitation of the AdS state to derive the entropy formula for dS.\footnote{Roughly, the entropy formula is obtained by counting the ways in which the excitations $N(L)$ can be distributed over the number of degrees of freedoms $C(L)$. The derivation is unfortunately only alluded to in \citet{verlinde_emergent_2017}. See however \citet[section 4.4]{VerlindeVisser} for a detailed, alternative derivation of the same formula.} This gives the same entropy as usually attributed to a region of dS-spacetime. The claim that dS is an excitation of AdS is thus in the end established as plausible by showing that a state counting method available from the AdS/CFT correspondence provides the usually accepted entropy formula for a dS spacetime region at its curvature radius $L$. 

A decisive element of establishing the dS-state as an excitation of the AdS-state in the energy-based argument strikingly consists of turning the inter-representational relation \textit{holding at the ground level between the AdS spacetime and a CFT} into an intra-theoretical one, that is reinterpret the number of configurations on the CFT side as in fact the number of configurations for realising the AdS spacetime (the AdS-state) \textit{simpliciter}. Thereby, a similar formula for the energy in terms of configurations as in the case of a CFT can be used at the excited level of the AdS state (where the correspondence to a CFT as such is normally not available anymore\footnote{Or at least not straightforwardly. See \citet{VerlindeVisser} for an attempt towards correspondence relations between non-AdS-spacetimes (such as de Sitter- and Minkowski-spacetime) and CFTs more generally. These specific correspondence relations are however based on connecting the non-AdS-spacetimes to AdS-spacetime, and thus again on the AdS/CFT correspondence in one way or the other.}). 
But also the entropy-based argument builds on re-interpreting the AdS/CFT correspondence as an intra-theoretical statement; after all, the AdS/CFT correspondence is used to single out a specific ground state corresponding to AdS --- the holographic state --- which then allows for reading dS as an excitation.

\subsection{Holography without holography}
We now argue that both \citet{verlinde_emergent_2017} and \citet{jacobson_entanglement_2016} in their respective ways employ insights gained from the AdS/CFT correspondence to advance non-holographic theories of quantum gravity; in accordance with the general tenet of the heuristic function of dualities as identified by \citet{de_haro_heuristic_2018}. We call this particular methodology `holography without holography'. 


In Jacobson's case, an entire derivation in the context of the AdS/CFT correspondence served as a guide towards an analogous intra-theoretical, and thus more promising, derivation in a non-holographic setting made possible by means of additional (plausible) assumptions. Verlinde uses the tensor network analogue of the AdS/CFT correspondence to model the degrees of freedom in the AdS vacuum (exploiting the holographic stabiliser condition implied by the AdS/CFT correspondence), and then speculates that this generalises even when the tensor network is no longer holographic (due to long range entanglement). This generalisation is then also supported by independent consistency checks against known results for de Sitter spacetime entropy scaling. In both cases, crucial aspects of the formal framework surrounding the relevant parts of the AdS/CFT correspondence are kept in place to secure the expedience in the non-holographic setting: In Verlinde's case the tensor network structure and in Jacobson's case non-holographic equivalents of crucial relations such as eq. \ref{eq:RyuTakayanagi} and eq. \ref{AB=EB}. The work already done in the context of the AdS/CFT correspondence ensures that these are controlled environments which enable the type of manipulation necessary for promoting the inter-representational relations of the AdS/CFT correspondence into intra-theoretical relations. 

We conjecture that the promises of this methodology is not idiosyncratic to the two cases studied but can be generalised to generic inter-representational relations (and structures of relations) in the AdS/CFT correspondence and possibly beyond. The overall heuristic of holography without holography can be decomposed into the following steps:
\begin{enumerate}
    \item Identify features of the AdS/CFT correspondence that are desirable, \textit{if} they can be conceived as intra-theoretical relations.
    \begin{description}
        \item[Jacobson:] The relation between EFEs and entanglement (entropy).
        \item[Verlinde:] Holographic stabiliser condition for entanglement networks.
    \end{description}
    \item Identify the specific formal (sub)framework of the AdS/CFT correspondence that sustains these features.
    \begin{description}
        \item[Jacobson:] The Ryu-Takayanagi formula, the relation between entanglement entropy and energy, and the relation between area and energy.
        \item[Verlinde:] Tensor network formalism.
    \end{description}
    \item Seek to embed this formal (sub)framework into the intra-theoretical context of interest modifying it adequately, while preserving enough of its structure such that it can still sustain the features of interest.
    \begin{description}
        \item[Jacobson:] Entanglement equilibrium, (leading order) area scaling of entanglement entropy, and (approximate) conformal invariance.
        \item[Verlinde:] Identify the holographic stabiliser condition for entanglement network as a ground state, thus allowing for (and requiring at the same time) its violation at the level of excited states.
    \end{description}
    \item Exploit the setup to derive \textit{intra-theoretical} results analogous to the inter-representational ones of the AdS/CFT correspondence and/or utilize the now intra-theoretic formal (sub)framework to inform the physics of an otherwise intractable context.
    \begin{description}
        \item[Jacobson:] Derives the Einstein Field Equations from entanglement constraints.
        \item[Verlinde:] Obtains excitation states (in particular, dS comes out as an excitation of AdS).
    \end{description}
\end{enumerate}

Given that holography without holography requires non-trivial innovative creativity in its application --- as signified in both case studies ---, it is by no means an automaton for discovery. But it suggests a way to activate the AdS/CFT correspondence in the context of discovery for general purposes of quantum gravity beyond the string theory on the AdS side.


\section{AdS$\stackrel{dual}{=}$CFT as a guiding principle to quantum gravity}

So far, we first defended the view that the AdS/CFT correspondence in itself does not break new grounds towards the quantum origin of gravity. We then went on to argue that it can still excel in a methodology referred to as `holography without holography'. Reconsidering the role of the AdS/CFT correspondence as an \textit{inspirational template} for intra-theoretical relations from inter-theoretical relations, the AdS/CFT correspondence got to be understood as the key guiding principle for applying the methodology of holography without holography.
In this section, we make our conception of the AdS/CFT correspondence as a guiding principle more precise. In particular, we explain that the AdS/CFT correspondence in the context of holography without holography is best conceived of as what we suggest calling an \textit{analytic} guiding principle.
\footnote{This is a specific way in which the AdS/CFT correpondence can be said to have a `heuristic function' (cf. \cite{de_haro_heuristic_2018}).}

It is not a secret that the search of a theory of quantum gravity suffers from the unavailability of empirical data. Therefore, in addition to efforts of extending quantum phenomenology---the search for new means of experimentally probing the relevant regimes in which we expect quantum gravitational effects to kick in---insights into a theory of quantum gravity can otherwise only come from the imposition of principles and the implementation of these principles within specific theoretical programs such as string theory, loop quantum gravity, or asymptotic safety (to mention a few). Towards a theory of quantum gravity, such principles provide guidance, motivation of the problem of quantum gravity in the first place\footnote{Without much empirical data at all, there is hardly any empirical data in need of explanation, either. Thus, the motivation for quantum gravity is largely theoretical; as such, it typically rests on the demand for realisation of certain principles such as UV completion (the idea that the theory holds formally up to all high energies) or unification.} and sorts of non-empirical justification\footnote{\citet{Dawid} for instance suggests adopting means of non-empirical theory \textit{confirmation}, at least in the context of string theory. In any case, it remains uncontroversial that, in the ongoing context of discovery, principles help in the preliminary appraisal of hypotheses and theory proposals at the level of plausibility arguments (see \citet{sep-scientific-discovery}, section 9.2).}. Examples of guiding principles include \textit{UV-completion}, and \textit{quantisation}; examples of (weakly) justificatory principles include \textit{minimal length}\footnote{Roughly, the idea that spatiotemporal structure consists of discrete chunks. See \citet{Hossenfelder:2012jw} for more.} and \textit{quantisation}; and examples of motivation include \textit{unification}, to name some.\footnote{See \citet{CrowtherLinnemann} for a detailed discussion.} Notably, such principles are intentionally not strictly formalised in order to keep them as framework-independent as possible (for instance, the idea of UV completion, that a theory holds up to arbitrarily high energies, can upon its general conception still be fleshed out in different specific scenarios). Now, we see it as a task for the philosopher to work out the principles used and alluded to in different approaches, and to make their mutual dependence relationships transparent (to give an example: in many frameworks, the principle of minimal length implies UV completion, raising the question whether minimal length is the intended principle one wants to commit to --- or rather UV completion). Our discussion of the AdS/CFT correspondence as a sort of guiding principle can be seen as a contribution to this more general project.\footnote{At the end of the day, the goal would be to create a hierarchy network of principles which provide a powerful tool for comparing the commitments in different approaches to one another, and allow for exploring further options upon dropping or adding the commitment to certain principles in specific approaches.}

Now, the duality aspect of the AdS/CFT correspondence entails a (conjectured) mathematical isomorphism whereas most guiding principles are in comparison (also) desiderata for instantiations of certain physical or metaphysical\footnote{Unification as a guiding principle for instance has both physical and metaphysical strands. For a discussion, see \citet{Salimkhani}.} properties (call these \textit{physical} and \textit{metaphysical guiding principles} respectively). This becomes clearer via examples: unification is usually understood (or desired) to express not just unity of representation but unity of \textit{nature}; UV completion not just (predictive) completeness of our description (our theory) but as predictive completeness in our picture of nature; etc. In light of this contrast, we refer to the AdS/CFT correspondence as a (merely) \textit{analytic guiding principle} as it is \textit{not} intended to apply to nature as such, that is to have any significance other than that of formal relationships.

To clarify the notion of an analytic guiding principle and the role of the AdS/CFT correspondence qua analytic guiding principle, it is instructive to make a comparison to other guiding principles which can be conceived of as analytic, such as quantisation and minimal coupling.\footnote{Both are applicable in contexts well beyond quantum gravity.}$^{\text{,}}$
\footnote{One might object to the classification of quantisation as an analytic guiding principle given that physical considerations are taken into account in quantisation (for instance, in order to avoid ambiguities, one can demand that certain classical symmetries must have corresponding quantum symmetries, that is quantum anomaly avoidance); however, quantisation theory as such covers a vast space of theories for which a clear physical interpretation is not readily given. It is for this reason that quantisation in our view first of all amounts to a mathematical (that is, formal) theory.}
Quantisation is a prescription for going over from the classical to the quantum theory. In some sense, it amounts to a (not always unambiguous)\footnote{Consider for instance the normal ordering issue. See for instance \citet{Landsmann}.} prescription taking classical observables to quantum observables as well as classical evolution to quantum evolution. There are different attempts to make this notion technically rigorous (as for instance geometric quantisation, and deformation quantisation); none of these can however provide a satisfactory account of quantisation in every desirable scenario (see for instance \citet{Landsmann}).\footnote{In other words, the general principle of quantisation can only be cashed out (so far) through a cluster of various (not completely mutually compatible) technical renderings, which is however the case for many other principles (such as unification, or UV completion) as well. This is not a problem: the notion of principle is after all supposed to capture a general idea at least \textit{prima facie} independent of framework-specific theoretical renderings.} 
Minimal coupling in the context of GR is a specific prescription for associating matter equations in flat spacetime to corresponding matter equations in curved spacetime.\footnote{At the level of dynamical equations, any instance of the flat metric is substituted by the corresponding general relativistic metric; and any instance of the flat spacetime covariant derivative by the covariant derivative associated with the general relativistic metric.} As a prescription, it suffers from ordering ambiguities in mapping higher derivative expressions in flat spacetime to that in curved spacetime, analogous to those occurring in quantisation (see \citet[chapter \S 16.3]{MisnerThorneWheeler}. Quantisation and minimal coupling are---just like the AdS/CFT correspondence---mathematical correspondence principles. Whereas both quantisation and minimal coupling can be straightforwardly understood as recipes (and thus methodologies) for theory change, in contrast, the AdS/CFT correspondence is a \textit{varying} ingredient for fleshing out the recipe of holography without holography (see section 4.3), and not a recipe for theory change itself:

\begin{itemize}
    \item 
    in the context of Jacobson's work, the inter-representational AdS/CFT correspondence serves as \textit{a formal template} towards corresponding intra-theoretical relations without being realised (not even approximately) at any point itself. The AdS/CFT correspondence is used as a formalistic (thus analytic) guide towards inter-theoretical relations in a specific framework chosen by Jaocobson while holography as a physical feature falls out of the picture.
   \item 
   in the context of Verlinde's work, the inter-representational AdS/CFT correspondence in its tensor network formulation is used to define the ground state of the system based on the stabiliser condition for holography; thus again embedding parts of its formal framework in an intra-theoretical context. Relying on established techniques from the tensor network formulation of the AdS/CFT correspondence, the nearest neighbour entanglement due to the stabilizer condition can be systematically broken to suggest possible excitations of the system. 
\end{itemize}

Holography without holography, in other words, forms a general strategy in the context of the AdS/CFT correspondence that can be carried out in different ways. The unifying idea is, as explicated in section 4.3, the identification of those inter-representational features of the AdS/CFT correspondence that are desirable if they can be made intra-theoretical, and the subsequent embedding of the formal framework sustaining these features into the intra-theoretical context of interest. What are desirable features will depend on one's purposes, and how to succeed with the embedding will depend on the specific context.\footnote{Again, as emphasized in section 4, this embedding of originally inter-representational elements into an intra-theoretical context cannot be achieved by simply stipulating that the relations of interest are not exact.}

A substantial point of working out the methodology behind holography without holography is to acknowledge an even more high-level tranformational methodology of \textit{changing an inter-representational into an intra-theoretical relation}. After all, a methodology analogous to that of holography without holography has for instance been tacitly assumed in discussions on the empirical relevance of global symmetries for longer now: global symmetries of say Newtonian theories (such as global translation/rotation/... invariance) are --- as such ---, pace orthodoxy, inter-representational relations.\footnote{Global symmetries are arguably analogous to dualities. To use the slogan mentioned by \citet{deHaroButterfield2017}, ``a duality is like a symmetry, but at the level of a theory" (p. 6). See paragraph (2) in section 1.1 therein and \citet{read_motivating_2018} for a more detailed discussion of the analogy.} They tell us how two different Newtonian \textit{representations} (two different Newtonian ``worlds") can be mapped into another. They however obtain empirical relevance (say become observable --- either directly, or indirectly) via associated conserved quantities --- as usually done\footnote{See for instance \cite{GreavesWallace}.} --- when understood as possible transformations of subsystems with respect to a background structure in \textit{one} Newtonian representation, that is as intra-theoretical relations.
So, an overarching methodology behind holography without holography includes changes from inter-representational duality relations into intra-theoretical relations (with adequate new theoretical embedding) like those by Jacobson and Verlinde; but also, for instance, (2) the change from an inter-model \textit{symmetry} to an intra-model (physical) symmetry. Note that the overarching methodology suggested through holography without holography is hereby still more specific than the generic heuristics of ``change one of your guiding principles" or ``change your guiding principle of holography".



\subsection{Objections and replies}
Nevertheless, what is the status of the \textit{specific} methodology of holography without holography employed in this paper, especially given that we are willing to build such bold (general) methodological claims on it? Two core objections suggest themselves: (1) The approaches considered are highly speculative and only followed by a small group of researchers; in other words, they are neither empirically nor sociologically well supported. Why then think that anything of general value can be concluded from their consideration? (2) Even in the extremely lucky scenario that the approaches of Jacobson or Verlinde turned out to culminate in anything close to empirically adequate theories, why should it matter how we got to them (if this methodology was used at all)? 

Let us start by addressing the second objection. First, it is not important whether Jacobson or Verlinde actually used our proposed methodology extracted from their approaches\footnote{That \citet[1]{jacobson_entanglement_2016} has used a methodology resembling holography without holography in the context of discovery is suggested by his clear reference in the paper to the work by van Raamsdonk and collaborators as a source of inspiration quoted above.} but that it could have been used.\footnote{Compare this for instance to the distinction between `discoverability'---something could have been discovered at a certain point of time---and `discovery'---something was actually discovered at a certain point of time---by \cite{nickles1988truth}. Arguably, discoverability is more interesting than actual discovery: working with the notion of discoverability is more likely to reveal stable conceptual relations while decreasing the risk of taking historical contingencies as relevant.} That a rationale is not used, does not rob the rationale its relevance.\footnote{See \cite{Nickles}.} Then, turning the objection around, one way to understand our work is in fact as (arguably another) \textit{demonstration} --- contra long-lived Popperian biases --- that there is a rationality of discovery\footnote{Dividing up the context of discovery into a context of generation, and a context of pursuit (following \cite{Laudan}), the statement is really that there is (some) rationality (not necessarily logic though) for generating scientific hypotheses and perhaps even theories. That there is a rationality in the context of pursuit (say in the form of appraising plausibility arguments to decide which research direction to tackle next), is much less debated (see \cite{nickles1988truth}).} possible in the context of emergent gravity and --- just as already hinted at in various other special cases\footnote{Consider for instance the work of \cite{Darden} on how specific biological mechanisms can be systematically arrived at through instantiating abstract mechanism schemes.} --- \textit{a fortiori} in science. 

Concerning the first objection we cannot do much more than stay defiant: In the end, it is up to us philosophers of physics whether we stick to grand plans of helping out physicists in the moment of crisis (as for instance called for by the \cite{huggett_emergent_2013}) or back off in the very first moment that we realise that ongoing scientific research can simply turn out to be ill-directed (as any fallible enterprise)). To engage in current research, whether as a physicist or as a philosopher (and at least scientists know this), also amounts to accepting the possible but unknown opportunity cost.
To address the sociological aspect of this objection specifically: not the number of individuals working in an approach but rather the groundedness in accepted principles and spontaneous reproduction of approaches by more or less independent researchers---as it is the case for the works of Jacobson and Verlinde---strike us as sensible criteria for which proposals to consider. Note also that in a sense it is not correct that Verlinde and Jacobson's approaches are just pursued by a few individuals; their programs are part of a general trend towards considerations of horison and entanglement entropy in order to make a step forward towards a theory of quantum gravity\footnote{See for instance works by \citet{maldacena_cool_2013},
\citet{chirco_spacetime_2014}, \citet{han_loop_2017}, \citet{baytas_gluing_2018}, \citet{cao_bulk_2018}, and \citet{chirco_group_2018}.}. Working out a (joint) rationality of discovery behind their approaches, should thus benefit a whole field.

\section{Conclusion}
In this paper we uncovered an enticing methodology where inter-representational relations are, at least for heuristic purposes, turned into intra-theoretical ones. Both of our two case studies focused on how in particular the AdS/CFT correspondence can be put to use in non-holographic settings when re-interpreted intra-theoretically. It was argued that this is of particular relevance in the context of the AdS/CFT correspondence: after all, although the AdS/CFT correspondence promises a relation between gravitational and quantum degrees of freedom, its nature as a duality prohibits it from being a theory of the quantum origin of gravity. While this complication has been alluded to by a number of authors, this paper offered more concrete details how to activate the AdS/CFT correspondence for general purposes in quantum gravity: the methodology of holography without holography. This methodology was concretely exemplified in the two case studies of works by \citet{jacobson_entanglement_2016} and \citet{verlinde_emergent_2017}, who in their respective ways draw inspiration from the (inter-representational) AdS/CFT correspondence for making an intra-theoretical claim. Thus, we argued that the AdS/CFT correspondence here serves as an analytic guiding principle---consisting of purely mathematical relations among representations of the bare theory---that can inform intra-theoretical relations for the purpose of developing prospective theories for the quantum origin of gravity.

In so far as holography without holography proves to be successful beyond these two case studies, it serves as an \textit{apology} for the extensive research on the AdS/CFT correspondence. Critics argue that such research is misguided since the actual world is not AdS; that research on the AdS/CFT correspondence is thus merely esoteric mathematics re-expressing a wrong-headed theory and, to put it boldly, the equivalent of studying the relation between bachelors and unmarried men in a world devoid of men. Holography without holography offers a response to such criticism: the AdS/CFT correspondence can serve as an important guiding principle towards a theory of quantum gravity for the actual world despite the fact that it is not realised (probably not even approximately). Given the current status of quantum gravity research we should focus more on the indirect contribution of the AdS/CFT correspondence as it transpires through the methodology of holography without holography.

\bibliography{bib.bib}
\end{document}